\documentclass[usenatbib,letters]{mn2e}

\newcommand\dosingle[1]{#1}  \newcommand\dodouble[1]{ } 

\newcommand\nice[1]{#1}    \newcommand\subm[1]{}   
\subm{ \renewcommand\dosingle[1]{ }   \renewcommand\dodouble[1]{#1} }

\dodouble{\documentclass[usenatbib,referee]{mn2e}}

\usepackage{graphicx}

\usepackage{amsfonts}

\usepackage{mn_bmath} 

\usepackage[varg]{txfonts}
\usepackage{fixltx2e} 

\newcommand\mystamp[1]{#1}

\newcommand\mystamppreamble{
  \usepackage{eso-pic}
  \usepackage{color} 
  \definecolor{redstamp}{rgb}{0.99,0.80,0.90} 
  \usepackage{datetime}
  \usepackage[normalem]{ulem}
}

\mystamp{\mystamppreamble}

\usepackage{natbib}  
\usepackage{url}

\newcommand\prerefereechanges[1]{#1}

\newcommand\postrefereechanges[1]{#1}

 \usepackage{hyperref}

\nice{ 
 \providecommand{\eprint}[1]{\href{http://arxiv.org/abs/#1}{{\tt [arXiv:#1]}}}

\providecommand{\url}[1]{\href{#1}{#1}}

}

\providecommand{\adsurl}[1]{} 
\providecommand\apj{ApJ}                 
\providecommand\apjs{ApJSupp}                 
\providecommand\aap{A\&A}            
\providecommand\mnras{MNRAS}

\providecommand\prd{PRD}

\providecommand\nat{Nat.}
\providecommand\grg{Gen. Rel. Grav.}
\providecommand\annalesBruxelles{Annales de la Soci\'et\'e Scientifique de Bruxelles}

\nice{  }

\newcommand\gtapprox{\,\lower.6ex\hbox{$\buildrel >\over \sim$} \, }
\newcommand\ltapprox{\,\lower.6ex\hbox{$\buildrel <\over \sim$} \, }
\newcommand\propapprox{\,\lower.6ex\hbox{$\buildrel \propto\over \sim$} \, }

\newcommand\arcs{\ifmmode {'' }\else $'' $\fi}     
\newcommand\arcm{\ifmmode {' }\else $' $\fi}       
\newcommand\ddeg{\ifmmode^\circ\else$^\circ$\fi}    

\newcommand\frtoday{Le\space\number\day\space\ifcase\month\or
  janvier\or f\'evrier\or mars\or avril\or mai\or juin\or
  juillet\or ao\^ut\or septembre\or octobre\or novembre\or 
d\'ecembre\fi\space \number\year}
\def\frdutoday{du\space\number\day\space\ifcase\month\or
  janvier\or f\'evrier\or mars\or avril\or mai\or juin\or
  juillet\or ao\^ut\or septembre\or octobre\or novembre\or 
d\'ecembre\fi\space \number\year}

\newcommand\todayISO{\number\year-\ifnum\month<10 0\fi\number\month-\ifnum\day<10 0\fi\number\day}

\newcommand{\CD}{{\cal D}}
\newcommand{\CE}{{\cal E}}

\newcommand{\CM}{{\cal M}}

\newcommand\cqg{ClassQuantGra}   %
\newcommand\hMpc{\mbox{$h^{-1}$ Mpc}}

\newcommand\Ommzero{\Omega_{\mathrm{m0}}}
\newcommand\OmLamzero{\Omega_{\Lambda0}} 

\begin{document}

\title[Is the BAO peak a standard ruler?]{Is the baryon acoustic oscillation peak a cosmological standard ruler?}

\author[Roukema et al.]{Boudewijn F. Roukema$^{1,2}$,
  Thomas Buchert$^{2,3}$,
  Hirokazu Fujii$^4$, 
  Jan J. Ostrowski$^{1,2}$
  \\
  $^1$ Toru\'n Centre for Astronomy, 
  Faculty of Physics, Astronomy and Informatics,
  Grudziadzka 5,
  Nicolaus Copernicus University,
  ul. Gagarina 11,\\87-100 Toru\'n, Poland 
  \\
  $^2$
  Universit\'e de Lyon, Observatoire de Lyon,
  Centre de Recherche Astrophysique de Lyon, CNRS UMR 5574: Universit\'e Lyon~1 and \'Ecole Normale\\ Sup\'erieure de Lyon, 
  9 avenue Charles Andr\'e, F--69230 Saint--Genis--Laval, France\protect\thanks{BFR: during visiting lectureship; JJO: during long-term visit.}\\
  $^{3}$Departamento de Astronom\'ia, Universidad de Chile, Camino del Observatorio 1515, Santiago, Chile\\
  $^{4}$Institute of Astronomy, University of Tokyo, 2-21-1 Osawa, Mitaka, Tokyo 181-0015, Japan}  


\date{\frtoday}



\newcommand\Nchainsmain{16}
\newcommand\Npergroup{four}

\maketitle

\begin{abstract}
In the standard model of cosmology, the Universe is static in comoving
coordinates; expansion occurs homogeneously and is represented by a
global scale factor. The baryon acoustic oscillation (BAO) peak
location is a statistical tracer that represents, in the standard
model, a fixed comoving-length standard ruler.  Recent gravitational
collapse should modify the metric, rendering the effective scale
factor, and thus the BAO standard ruler, spatially inhomogeneous.
Using the Sloan Digital Sky Survey, we show to high significance
($P<0.001$) that the spatial compression of the BAO peak location
increases as the spatial paths' overlap with superclusters increases.
Detailed observational and theoretical calibration of this BAO peak
location environment dependence will be needed when interpreting the
next decade's cosmological surveys.
\end{abstract}


\mystamp{}


\dodouble{ \clearpage } 


\newcommand\FIGWIDTHONE{1.05\columnwidth}
\newcommand\FIGWIDTHTWO{1.05\columnwidth}

\newcommand\fcompression{
  \begin{figure}
    \includegraphics[width=\FIGWIDTHONE]{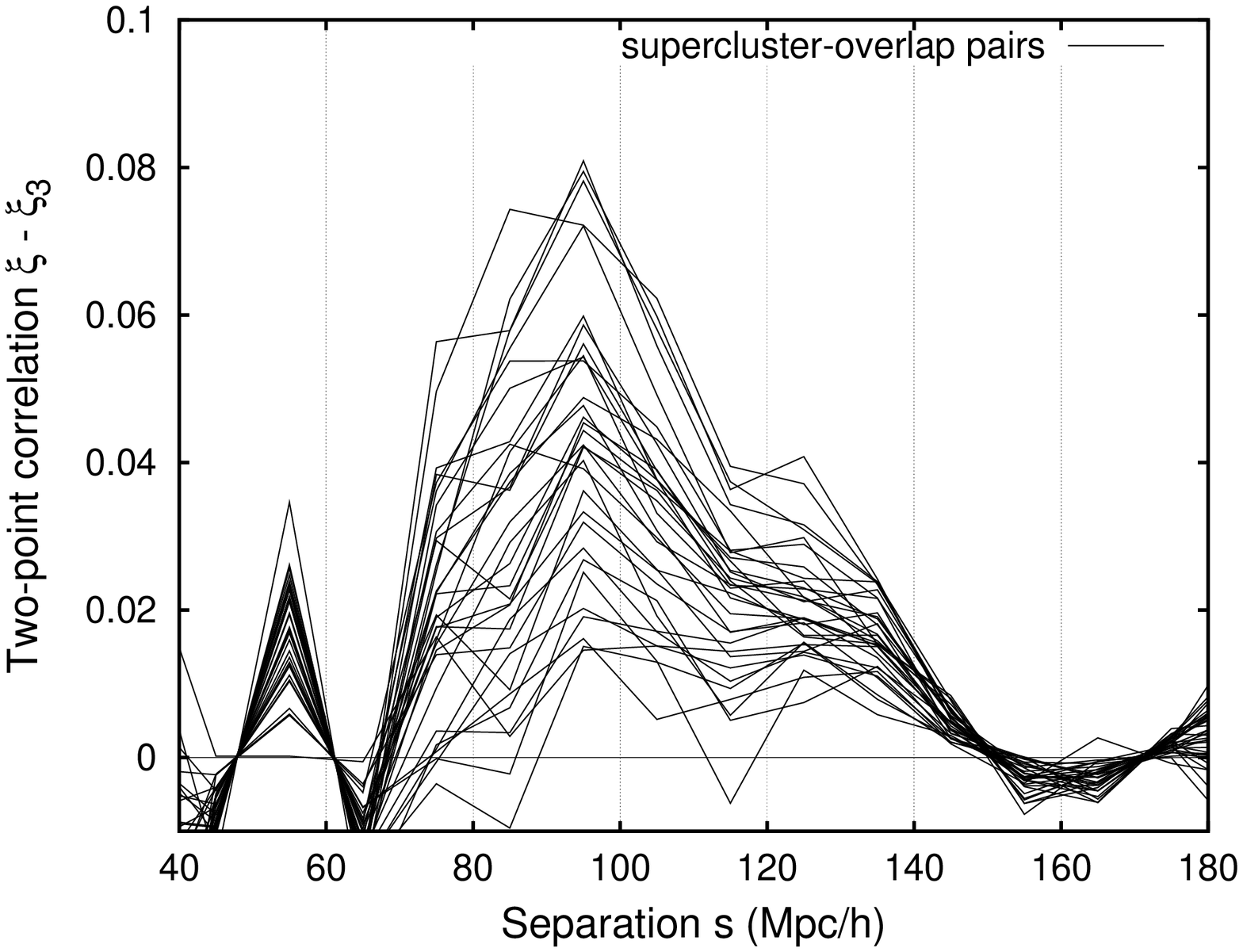}
    \includegraphics[width=\FIGWIDTHONE]{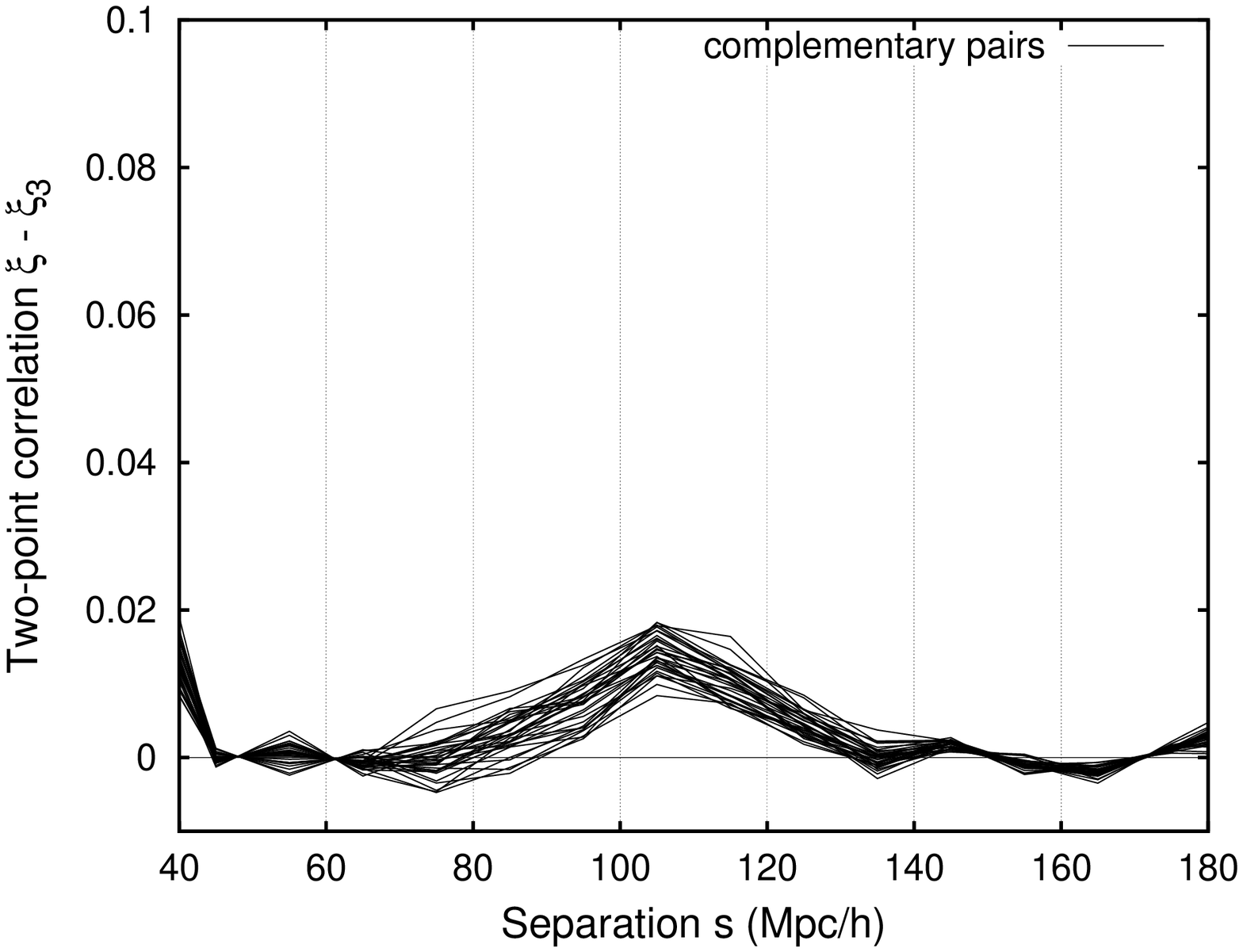}
    \caption{{Compression of the baryon acoustic oscillation peak.}
      \protect\prerefereechanges{{\em Upper panel:}} BAO peak (cubic-subtracted correlation function) for
      pairs of luminous red galaxies whose paths either overlap with
      superclusters by $\omega \ge \omega_{\min} =60${\hMpc} or are
      entirely contained within the superclusters.
      {The overlap $\omega$ is the chord
        length defined \protect\citep[][Sect.~2.3, Fig.~1]{RBOF15} by the
        intersection of the path joining two LRGs and the
        supercluster modelled as a sphere.}
      Individual curves represent 32 
      bootstrap resamplings of the supercluster 
      catalogue (235 objects) and the 
      ``random'' galaxies (484,352 selected from 1,521,736). 
      The real galaxies (30,272) are not resampled.
      Most of the curves peak
      sharply at 95{\hMpc}; a few peak at 85{\hMpc}. 
      The high amplitude (in comparison with 
      \protect\prerefereechanges{the lower panel}) is
      consistent with biasing that modifies the amplitude
      of $\xi$.
      \protect\prerefereechanges{{\em Lower panel:}}
      BAO peak for the complementary subset of galaxy
      pairs. The peak occurs at the standard value of about
      105{\hMpc}.
      \label{f-thr60}}
\end{figure}} 

\newcommand\foverlapdependence{
\begin{figure}
  \includegraphics[width=\FIGWIDTHTWO]{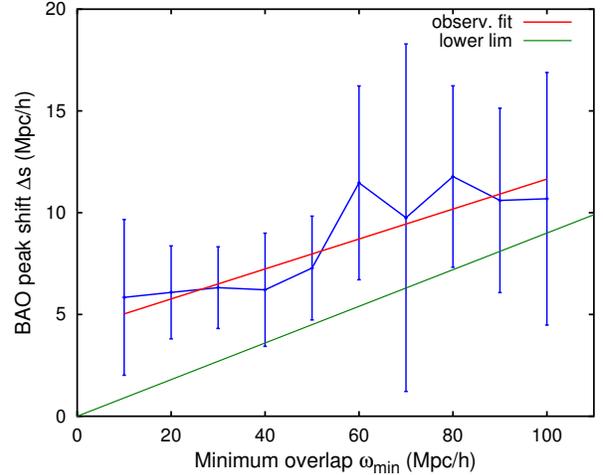}
  \caption{{Overlap dependence of the BAO peak
      shift.}  BAO peak shift $\Delta s :=
    s\textsubscript{non-sc} - s_{\mathrm{sc}}$, where
    $s_{\mathrm{sc}}$ and $s\textsubscript{non-sc}$ are the
    median estimates of the centres of the best-fit
    Gaussians to $\xi-\xi_3$ for LRG pairs that overlap
    superclusters (sc) and those that do not (non-sc),
    respectively. The error bars show a robust estimate
    of the standard deviation, $\sigma(\Delta s)$,
    defined here as 1.4826 times the median absolute
    deviation of $\Delta s$. At each $\omega_{\min}$,
    these statistics are calculated over 32 bootstrap
    resamplings of the observational
    {data.}  
    {Four out of the 320 
      Gaussian fits in the {supercluster-overlap}
      case failed
      and were ignored in calculating these statistics; no failures
      occurred for the 320 non--supercluster-overlap cases.}
    A linear
    least-squares best fit relation 
    {$\Delta s =
      4.3${\hMpc}$+ 0.07 \omega_{\min}$} is shown
    \protect\postrefereechanges{with a red line.}
    The most
    significant individual rejection of a zero shift is
    {$\Delta s = (6.3\pm2.0)${\hMpc} for
    $\omega_{\min}=30${\hMpc}, a 3.1$\sigma$} (Gaussian)
    rejection. {Since
      bootstraps are used, this estimate is conservative:
      $\sigma(\Delta s)$ is expected to be an
      overestimate of the true
      uncertainty \protect\citep[][Sect
        2.2]{FisherIRAS94bootupper}.}
    \protect\postrefereechanges{A 9\% shift
      [Eq.~(\protect\ref{e-BAO-RZA-first-estimate})]
      would give
      $\Delta s = 0.09 \omega$. Since $\omega \ge \omega_{\min}$,
      a scalar-averaging lower expected limit
      $\Delta s > 0.09 \omega_{\min}$ is shown as a green line.}
    \label{f-overlap-dependence}}
\end{figure}} 

\section{\protect\prerefereechanges{Introduction}}

The choice of a spacetime coordinate system for the
Universe that enables the expansion to be represented
via {a global} 
scale factor $a$ as a function of just one
coordinate (cosmological time $t$, \citealt[e.g.][]{Lemaitre31ell}) is extremely
convenient.  On large enough comoving length scales,
statistical spatial patterns are fixed in this
coordinate system.  Thus, the theory of primordial
density fluctuations leads to that of baryon acoustic
oscillations (\citealt{EisenHu98Tofk}; BAOs).  The BAO peak
in the two-point spatial auto-correlation function
$\xi$ was clearly detected in the Sloan Digital Sky
Survey \citep{Eisenstein05} and the Two-Degree Field
Galaxy Redshift Survey (\citealt{Cole05BAO}; earlier surveys
may have detected this too, \citealt{Einasto2Nat97}).  BAOs
now {constitute
  one of the most
  important tools for making cosmological geometry
  measurements, especially for upcoming 
  {observational projects}
  such as the space mission
  Euclid \citep{EuclidScienceBook2010} and the ground-based
  instruments
  {DESI \citep{Levi13DESI},}
  4MOST \citep{deJong12VISTA4MOST},
  and the LSST \citep{TysonLSST03}}. 
{The
BAO peak location} of about 105{\hMpc}
(where $h$ is the Hubble constant $H_0$ expressed in
units of 100 km/s/Mpc) {is} 
{commonly}
expected to be a large
enough comoving length scale for it to provide a fixed
comoving ruler in the real Universe.

\fcompression

\prerefereechanges{However, the validity of the BAO peak location
as a standard ruler depends on the validity of the assumed
cosmological metric 
\prerefereechanges{(differential rule for measuring lengths)}
in the context of the real Universe, which
is lumpy \citep[cf.\/ the ``fitting problem'',][]{EllisStoeger87}.}
Scalar averaging is a general-relativistic formalism
that extends beyond 
the standard cosmological model, by
allowing a spatial section of the Universe at a given
time to have an inhomogeneous metric and calculating
\postrefereechanges{background-free}
volume-weighted averages of scalar
variables 
\prerefereechanges{\citep{Buch01scalav,Buchert08status}}.  The univariate scale
factor $a(t)$ is replaced by an effective,
environment-dependent {volume-based} scale factor {$a_{\CD}(t) \propto V_{\CD}^{1/3}$,}
dependent on both the choice of compact spatial domain
$\CD$ {of volume $V_{\CD}$}
{and} on time.
Without this extension, the
\postrefereechanges{cosmological, comoving}
metric is forced
\postrefereechanges{(by definition)}
to be rigid in comoving
coordinates,
i.e. inhomogeneities in the matter
distribution are not allowed to ``tell \postrefereechanges{comoving} space how to
curve''
\postrefereechanges{\citep[e.g.][]{BuchCarf08}}.
Applying scalar averaging to an
observationally standard power spectrum that
statistically represents density fluctuations at an
early epoch implies that even for spatial domains as
large as the BAO peak length scale, the
{environment dependence} of the scale factor should be
observationally detectable, 
\postrefereechanges{i.e.} $a_{\CM} < a_{\CE}$ is expected, where $\CM$ 
(``Massive'') and $\CE$ {(``Empty'')}
represent overdense and underdense 
spatial regions, {respectively}.
\postrefereechanges{For example, adopting 1$\sigma$ initially overdense ($\CM$) and
  underdense ($\CE$) fluctuations in a spherical domain of diameter
  $\approx 105${\hMpc}
  and using equations (2), (13), (32), (50), and (54) of
  \citet*{BuchRZA2} to integrate
  the Raychaudhuri equation (9)
  of the same paper gives a relativistic Zel'dovich approximation
  estimate of
  \begin{equation}
    a_{\CM}/a_{\CE} \approx 0.91\;.
    \label{e-BAO-RZA-first-estimate}
  \end{equation}
  In other words, in the scalar averaging approach,
  one way in which matter inhomogeneities are expected
  to affect the large-scale
  geometry and dynamics is to shrink the curved-space
  volume of overdensities on the BAO scale by about
  $1 - (a_{\CM}/a_{\CE})^3 \approx 24\%$,
  leading to an expected shift in the BAO peak location
  to a lower scale
  by somewhat below or above 9\% for galaxy pairs that
  weakly or strongly, respectively, overlap with 
  typical BAO-scale overdensities.}

\postrefereechanges{An environment-dependent}
effect has recently been detected as a six percent
compression of the BAO peak location for spatial paths
that touch or overlap superclusters of luminous red
galaxies (LRGs) in the Sloan Digital Sky
Survey (\citealt{RBOF15}; SDSS; environment dependence of
$\xi$ at smaller scales has also been detected in the
SDSS, \citealt{Chiang15xidependsx}).

\foverlapdependence

\prerefereechanges{In this {\em Letter},}
we check whether the compression is dependent on
the minimum overlap between spatial paths and
superclusters, as it should be if the effect is induced
by the statistically overdense nature of the
superclusters. 

\section{\protect\prerefereechanges{Method}}
{We modify the previous method \citep[][Sect
  2]{RBOF15} in order to allow} 
stronger overlaps. 
{As in the original method, we calculate}
the correlation function $\xi$ of the ``bright'' sample
of LRGs in the SDSS Data Release~7 (DR7) for pairs of
LRGs selected for overlap \citep[][Sect.~2.3, Fig.~1]{RBOF15}
(or non-overlap) of
{superclusters} in the
survey \citep{NadHot2013}, using the Landy \&
Szalay estimator \citep{LandySz93} on real and ``random'' (artificial)
catalogues \citep{Kazin2010}. 
Comoving separations $s$
are calculated assuming the standard $\Lambda$CDM
{model \citep{WMAPSpergel,PlanckXVIcosmoparam13}}
with
matter density parameter $\Ommzero=0.32$ and dark
energy parameter $\OmLamzero = 0.68$.  A best-fit cubic
$\xi_3(s)$ over separations $s \le 70${\hMpc} and $s
\ge 140${\hMpc} (i.e. excluding the peak), is found for
the tangential signal (pairs $\le 45^\circ$ from the
sky plane). The procedure is repeated, bootstrap resampling the supercluster
catalogue \citep{NadHot2013} and the ``random'' LRG
catalogue, several times. The BAO peak location is estimated from the
medians and median absolute deviations of the centres
of the best-fit Gaussians to $\xi(s)-\xi_3(s)$. In
this work, minimum overlaps $\omega_{\min}$ in the
range $10${\hMpc} $\le \omega_{\min} \le 100${\hMpc}
rather than $\omega_{\min} = 1${\hMpc} are considered.
In order that $\xi$ be defined for $s \le
\omega_{\min}$ for these high values of
$\omega_{\min}$, we consider a pair of LRGs joined by a
comoving spatial path entirely contained within a
supercluster to satisfy the overlap criterion.

\section{\protect\prerefereechanges{Results}}

Figure~\ref{f-thr60} shows that for a minimum overlap
$\omega_{\min} = 60${\hMpc}, the BAO peak is shifted to lower
separations $s$ (panel a) than for the complementary set of LRG pairs (panel b).
The shift is clearer than for the earlier analysis, which had
$\omega_{\min} = 1${\hMpc} \citep[][Fig.~8]{RBOF15}. Requiring a 
stronger overlap yields a stronger shift.

Figure~\ref{f-overlap-dependence} shows the dependence of
the shift $\Delta s$ on $\omega_{\min}$. The BAO peak
shift for supercluster-overlapping LRG pairs appears to
increase from $\Delta s \approx$~6--7{\hMpc} for
$\omega_{\min} \ltapprox 50${\hMpc} to 
{$\Delta s
\approx$11{\hMpc}} for greater overlaps.  The Pearson
product-moment correlation coefficient of $\Delta s$ and
$\omega_{\min}$ is 0.87, with a probability of $P \approx
0.0008$, i.e. a positive correlation is detected to high
significance. Thus, the
{environment dependence} of the BAO
peak shift is confirmed.  Unfortunately, the
uncertainties shown in Fig.~\ref{f-overlap-dependence}
are too high to infer the details of this correlation
from the present data and analysis. The slope and
zeropoint of the linear best fit are 
{$0.073 \pm 0.040$
and $4.3 \pm 2.0${\hMpc},} respectively.

\postrefereechanges{As a rough guide to what is expected from scalar
  averaging, we can use the 9\% shift estimate from
  Eq.~(\protect\ref{e-BAO-RZA-first-estimate}), which 
  would give
  $\Delta s = 0.09 \omega$, where the overlap path lengths
  are approximated as corresponding to 1$\sigma$ overdense
  regions on the BAO scale, even though in reality,
  the overdense regions are superclusters, some smaller and some
  larger than this scale.
  Since $\omega \ge \omega_{\min}$,
  this implies a rough scalar-averaging lower expected limit of
  $\Delta s > 0.09 \omega_{\min}$, shown as a green line
  in Fig.~\ref{f-overlap-dependence}.}

\section{\protect\prerefereechanges{Discussion}}
Since the BAO peak location 
{serves 
as a major tool for cosmological geometry
measurements,}
it is clear that its {environment}
dependence will need to be observationally calibrated and
correctly modelled theoretically. 
It is possible that the effect could
  also be interpreted within the standard $\Lambda$CDM 
{model, as is the case for many
  large-scale phenomena. For example,}
{observed supervoids on the 200--300{\hMpc}
  scale \citep{NadHot2013,Szapudi14coldspotsupervoid} 
  {can be}
  interpreted within the $\Lambda$CDM model \citep{HotNad15ISWLCDMOK},
  although their occurrence is expected to be
  rare \citep{Szapudi14coldspotsupervoid}.  
  {In contrast, the study of SDSS DR7
    ``dim'' (or ``bright'') LRGs via Minkowski functionals 
    on scales ranging up to the BAO peak scale,
    within a 500{\hMpc} (or 700{\hMpc}, respectively) 
    diameter region, shows 3--5.5$\sigma$ (or 0.5--2.5$\sigma$)
    inconsistencies with $\Lambda$CDM
    simulations \citep*[][Table~1]{WiegBO14}.}
  Minkowski functionals have more statistical power than lower order
  statistics 
  {that are commonly used in 
    analysis of large-scale structure,}
  such as the two- and three-point correlation functions,
  the {correlation dimension} or percolation
  (``friends-of-friends'') 
  {analyses. This is because} all the $n$-point 
  correlation functions would be needed in order to represent the 
  statistical geometrical information that the Minkowski
  functionals contain.}\sloppy{}

\postrefereechanges{A possible avenue to studying the environment-dependent BAO shift within
  the standard approach would be to carry out a Fourier analysis rather
  than using the two-point correlation function. This would require the development
  of a supercluster-overlap--dependent Fourier analysis method.
  Another alternative, to avoid having to determine
  the position of the peak itself, would be comparison of radial to
  tangential correlation functions directly. This would require
  correcting for peculiar velocity effects, which are highly
  anisotropic with respect to the observer.}

Interpreting the environment dependence
of the BAO peak location
\postrefereechanges{reported in this {\em Letter}}
within the {standard} $\Lambda$CDM model
would require the comoving length scale at
which the Universe is rigid in comoving coordinates
to be pushed up to a scale greater than
105{\hMpc}.\sloppy{}
The environment dependence (e.g. the $\Delta s(\omega_{\min})$
relation) would have
to be modelled within a rigid comoving background that 
{can only exist} at {larger scales,} 
at which no sharp statistical feature 
{that can function as a standard ruler}
is presently known. This leads to 
a {Mach's principle} 
type of concern that at recent
epochs, it is difficult to have confidence that the standard
comoving coordinate system is correctly attached to an
observational extragalactic 
catalogue (peculiar velocity flow analyses indicate
similar concerns, \citealt{Wiltshire12Hflow}).
Interpretation within the scalar averaging
approach should be easier
{because its description
  of fluctuation properties and the cosmological expansion rate
  is environment-dependent.}

{Nevertheless,}
within the rigid comoving background framework (i.e. the standard model), 
{small shifts in the BAO peak location
  have been predicted analytically and from $N$-body 
  simulations \citep{Desjacques10BAObias,SherwZald12peakscale},
  while}
BAO reconstruction
techniques
\postrefereechanges{\citep{PadmanNWhite09reconstruct,PadmanN12reconstruct,Schmittfull15BAOrecon}}
have been developed to attempt
to evolve galaxies' positions backwards in cosmological time,
using a blend of theoretical calculations and $N$-body
models. The expected mean shift in the BAO peak location is
less than one percent, i.e. an order of 
magnitude less than what we find for the shift conditioned 
on $\omega_{\min} \ge 60${\hMpc}.
\postrefereechanges{The amplitude of the shifts found in these
  calculations is constrained by the assumption that curvature
  averages out on the assumed background, i.e. that a conservation
  law for instrinsic curvature holds globally
  \citep{BuchCarf08}.}

Theoretical work \postrefereechanges{is} underway in the
  scalar averaging approach, which general-relativistically extends the
  standard model,
  \postrefereechanges{allowing
    the restrictive assumption of a conservation law
    for intrinsic curvature to be dropped
    \citep{BuchCarf08}.
    The physical origin of curvature deviations from the background on scales as 
  large as the BAO scale can then be thought of 
  as following from the non-existence of a conservation law for 
  intrinsic curvature.}
To reconstruct the primordial comoving galaxy positions
more accurately {than in the standard model, i.e. 
to allow} flexible comoving curvature
that varies with the matter density and 
the extrinsic curvature tensor
{across a spatial slice},
{relativistic Lagrangian perturbation theory} 
(\citealt*{BuchRZA1,BuchRZA2}; \citealt{BuchRZA3})
is available for analytically guided calculations. 
$N$-body simulations in which the growth of inhomogeneities
is matched by inhomogeneous metric evolution will most likely
also
be needed to develop numerical confidence in what could be
called ``relativistic BAO reconstruction''.

\section{\protect\prerefereechanges{Conclusion}}
{No matter which approach is chosen,
  analytical, numerical and observational work will be
  required if the BAO peak location is to correctly function
  as a standard ruler for cosmological geometrical
  measurements, since the evidence is strong
  ($P<0.001$) that it is strongly affected by structure formation.
  Moreover, the formation of superclusters---in reality,
  filamentary and spiderlike
  distributions of galaxies \citep{Einasto14filspider}
  rather than the spherically symmetric objects assumed
  here for calculational speed---can now be
  tied directly to a sharp statistical feature
  of the primordial pattern of density
  perturbations.}

\section*{Acknowledgments}
  {Thank you to Mitsuru Kokubo for useful comments.}
  {The work of T.B. was conducted
    within the ``Lyon Institute of Origins'' under grant
    ANR-10-LABX-66.} 
  {T.B. acknowledges financial support from
    CONICYT Anillo Project (ACT--1122) and UMI--FCA (Laboratoire
    Franco-Chilien d'Astronomie, UMI 3386, CNRS/INSU, France, and
    Universidad de Chile) during a lecturing visit.}
  A part of this project was funded
  by the National Science Centre, Poland, under grant
  2014/13/B/ST9/00845. 
  {J.J.O. acknowledges support for part of this
    work from a National Science Centre, Poland, Etiuda 2 grant.}
  Part of this work consists of
  research conducted within the scope of the HECOLS
  International Associated Laboratory, supported in part
  by the Polish NCN grant DEC-2013/08/M/ST9/00664. A part
  of this project has made use of computations made under
  grant 197 of the Pozna\'n Supercomputing and Networking
  Center (PSNC). 
  Funding for the SDSS and SDSS-II has
  been provided by the Alfred P. Sloan Foundation, the
  Participating Institutions, the National Science
  Foundation, the U.S. Department of Energy, the National
  Aeronautics and Space Administration, the Japanese
  Monbukagakusho, the Max Planck Society, and the Higher
  Education Funding Council for England. The SDSS Web
  Site is
  \url{http://www.sdss.org}.\sloppy{}
  We gratefully acknowledge use of the 
  \citet{Kazin2010} version of SDSS DR7 real and random
  galaxies {at}
  \url{http://cosmo.nyu.edu/~eak306/SDSS-LRG.html}
  and of v11.11.13 of the \citet{NadHot2013}
  supercluster catalogue {at}
  \href{http://research.hip.fi/user/nadathur/download/dr7catalogue/}{{\tt http://research.hip.fi/user/}}
  \href{http://research.hip.fi/user/nadathur/download/dr7catalogue/}{{\tt nadathur/download/dr7catalogue}}.

\subm{ \clearpage }

%



\end{document}